\documentclass[journal]{IEEEtran} % use the `journal` option for ITherm conference style
\IEEEoverridecommandlockouts
\usepackage{cite}
\usepackage{amsmath,amssymb,amsfonts}
\usepackage{caption}
\usepackage{subcaption}
\usepackage{graphicx}
\usepackage{textcomp}
\usepackage{xcolor}
\usepackage{mwe}
\usepackage[ruled,vlined]{algorithm2e}
\usepackage{algpseudocode}
\def\BibTeX{{\rm B\kern-.05em{\sc i\kern-.025em b}\kern-.08em
    T\kern-.1667em\lower.7ex\hbox{E}\kern-.125emX}}
\begin{document}

\title{Physics-Informed Convolutional Autoencoder for Cyber Anomaly Detection in Power Distribution Grids

% {\footnotesize \textsuperscript{}}
 }
\author{{Mehdi Jabbari Zideh, \IEEEmembership{Student Member, IEEE},
Sarika Khushalani Solanki, \IEEEmembership{Senior Member, IEEE}}
\thanks{Authors are with the Lane Department of Computer Science and Electrical Engineering, West Virginia University, Morgantown, WV 26505 USA (e-mail: mj00021@mix.wvu.edu, skhushalanisolanki@mail.wvu.edu)}

% \thanks{Corresponding author: Mehdi Jabbari Zideh (e-mail: mj00021@mix.wvu.edu).}}

% \author{%%%% author names
%     \IEEEauthorblockN{1\textsuperscript{st} Mehdi Jabbari Zideh}% first author
%     , \IEEEauthorblockN{3\textsuperscript{rd} Sarika Khushalani Solanki}% delete this line if not needed
    % duplicate the line above as many times as needed to list all authors
   % , \IEEEauthorblockN{4\textsuperscript{th} Jignesh Solanki}% delete this line if not needed
   
    % \\%%%% author affiliations
    % \IEEEauthorblockA{\textit{Lane Department of Computer Science and Electrical Engineering, West Virginia University, Morgantown, USA}}\\% first affiliation

    % duplicate the line above as many times as needed to list all affiliations
    %%%% corresponding author contact details
    % \IEEEauthorblockA{email address or ORCID of corresponding author(s)}
}
\maketitle
\begin{abstract}
The growing trend toward the modernization of power distribution systems has facilitated the installation of advanced measurement units and promotion of the cyber communication systems. However, these infrastructures are still prone to stealth cyber attacks. The existing data-driven anomaly detection methods suffer from a lack of knowledge about the system's physics, lack of interpretability, and scalability issues hindering their practical applications in real-world scenarios. To address these concerns, physics-informed neural networks (PINNs) were introduced. This paper proposes a multivariate physics-informed convolutional autoencoder (PIConvAE) to detect stealthy cyber-attacks in power distribution grids. The proposed model integrates the physical principles into the loss function of the neural network by applying Kirchhoff's law. Simulations are performed on the modified IEEE 13-bus and 123-bus systems using OpenDSS software to validate the efficacy of the proposed model for stealth attacks. The numerical results prove the superior performance of the proposed PIConvAE in three aspects: a) it provides more accurate results compared to the data-driven ConvAE model, b) it requires less training time to converge c) the model excels in effectively detecting a wide range of attack magnitudes making it powerful in detecting stealth attacks.
\end{abstract}


\section{Introduction}\label{itroduction}
The recent advancement in smart grid monitoring technology has introduced the extensive installation of phasor measurement units (PMUs) and $\mu$PMUs to increase the situational awareness of power system operators \cite{situational_emma, pmu_emma}. This improvement in the system observability has exposed smart grids to various types of cyber-attacks threatening the grids' cyber-physical infrastructures \cite{reachability_FDIA}. As the behavior and patterns of the cyber anomalies are unknown and unpredictable, their detection is a challenging task to ensure secure and reliable operation of the system \cite{fdia_challenge}.

Machine learning methods as powerful techniques are applied to various aspects of power systems including forecasting \cite{solar_radiation}, power market operation \cite{two-sided, iterative}, power flow calculations \cite{physics_pf}, and cyber-physical security analysis \cite{CPS_ML}. More specifically, in the task of anomaly detection, several research investigations have applied deep neural networks (DNNs) to distinguish abnormal data from the normal data patterns within the sensor data. Authors in \cite{wavelet_transform} applied discrete wavelet transform (DWT) and a recurrent neural network (RNN) to detect cyber attacks. A similar methodology was proposed in \cite{DWT_HIF} to detect and classify high-impedance faults in power grids. The work in \cite{unsupervised_aligholian} proposed a generative adversarial network (GAN) model for event detection in $\mu$PMU measurements. 
A multivariate long short-term memory (LSTM) 
autoencoder-based model was employed in \cite{cyber_rafy} for cyber anomaly detection in power distribution systems.
% A Kalman filter and RNN model were presented in \cite{KFRNN} for the detection of false data injection attacks (FDIAs). 
% Although the data-driven ML methods are powerful in extracting features from vast amounts of data and can overcome the limitations of model-based methods, there are still some challenges including transferability, interpretability, and reliability problems in these methods.

Despite their impressive performance, ML models mainly suffer from poor extrapolation capabilities, scalability constraints, and lack of understanding of the system's physical laws. Consequently, this reduces their reliability as the system operators often struggle to interpret their implications. To address the aforementioned challenges and enhance the performance of the learning algorithms with lower computational effort, physics-informed neural networks (PINNs) were introduced integrating the fundamental physical principles of the systems into NN models \cite{raissi}. The works in \cite{PIML} and \cite{PINN} extensively reviewed the applications of PINN methods in the cyber-power domain and power systems, respectively. Recently a growing number of research studies have applied PINN methods for anomaly detection in power systems. Authors in \cite{PI_wind_farm} applied the physics-based inequality constraints to the isolation forest algorithm to detect falsification attacks on wind generation signals. The work in \cite{graph_ahmed} proposed a combined autoencoder with a graph neural network (GNN) model which learns the spatial structure of $\mu$PMU data through a weighted adjacency matrix to detect events in power distribution grids. A new framework for combining model-based and NN methods is proposed in \cite{VAE_LSTM_SE} where the prediction residuals of variational autoencoder (VAE), LSTM, and state estimation are fed to a DNN model for attack detection. A physics-informed method is employed in \cite{GECCN} for attack detection utilizing the information of transmission system topology structure and node features in graph edge-conditioned convolutional networks (GECCN). In \cite{data-driven_iot}, the physical relationships are integrated into the NN loss function through dynamical swing equations for the detection of load-altering attacks. 

In this paper, we propose a novel PINN model for cyber-anomaly detection in power distribution systems. More specifically, the contributions of this work are threefold: 1) We present a physics-informed convolutional autoencoder (PIConvAE) model by encoding the physics-based equations of $\mu$PMU measurements into the loss function of the model to detect cyber attacks; 2) The proposed model works in an unsupervised manner and does not require any prior labeling of the cyber-attacks; and 3) The efficacy of the proposed PIConvAE model is evaluated for cyber-attack detection on IEEE 13-bus and 123-bus systems.

The remainder of this paper is organized as follows. Section \ref{Attack_description} introduces data falsification attack mechanisms and data falsification functions. Section \ref{PINN_model} covers the principles and structure of the PINN model. Section \ref{Results} discusses the simulation results of the proposed PIConvAE model under falsification attacks. Finally, Section \ref{conclusion} concludes the paper.

\section{Development of Cyber-attacks on Power Distribution Measurements}\label{Attack_description}
The real-time data measured by $\mu$PMU and supervisory control and data acquisition (SCADA) systems including system frequency, voltage and current magnitudes and angles, are essential for power system control and stability analysis. The data transferred by the communication channels is prone to falsified attacks by the attackers to tackle data confidentiality, system reliability, and integrity. Assume that the measured data by the measurement units at time slot $t$ is $Z^i_{act}(t)$. The reported measurement $Z^i_{rep}(t)$ at time slot $t$ is equal to the measured data when the data is not falsified. However, if the data from measurement unit $i$ is compromised by an adversary, $Z^i_{act}(t)\neq Z^i_{rep}(t)$. We assume that the adversary can compromise the voltage and current magnitudes and angles. We apply the attack strategy employed in \cite{attack_context} where the falsified data is defined as  
\begin{equation}\label{additive_Eq}
Z^i_{comp}(t) = (1+\alpha (t)) Z^i_{act}(t) 
\end{equation}
where $\alpha (t)$ ($\alpha_{min} \leq\alpha (t)\leq \alpha_{max}$) describes the attack magnitude at time slot $t$, $\alpha (t)>0$ for additive attacks and $\alpha (t)<0$ for deductive attacks. For combined attacks, the compromised data consists of additive and deductive attacks in equal proportions. The objective of this type of attack is to report the falsified data as normal even when the data is above or below a certain threshold. Therefore, the control systems are deceived to not function properly. In this work, various values for $\alpha$ ($-5\%\leq\alpha(t)\leq5\%$) are chosen to not only demonstrate the variability of attack intensities but also to make them less readily detectable.

\section{Physics-informed Neural Network for FDIA Detection} \label{PINN_model}
In this section, we propose the physics-informed data-driven method by integrating the physics-based relationships of $\mu$PMU measurements into the ConvAE model to achieve more accurate and robust detection of falsified attacks in power distribution grids. 
% \subsection{Problem Statement}\label{problem_statement}
Suppose that we have a system with $N$ buses comprising distributed energy resources (DERs) such as PV and wind generation units, and power loads. Assuming the local measurements of the buses by measurement units, the following quantities are measured for each bus $i$
\begin{equation}\label{input}
 x_i = [ V_i , I_i, \theta_{i}, \delta_{i}, P_i, Q_i]
\end{equation}
where $V_i$ and $I_i$ are voltage and current magnitudes, $\theta_i$ and $\delta_i$ are voltage and current phase angles, and $P_i$ and $Q_i$ are active and reactive power measurements.

\subsection{Autoencoder-based Attack Detector}\label{Autoencoder}
Attack detection is a challenging task due to the unpredictable nature and patterns of attacks and their evolution over time. This motivates the applications of unsupervised learning methods, which can distinguish abnormal data from normal data samples without relying on data labels. Autoencoders (AE) as a category of unsupervised neural networks are trained on only normal training data and identify data that deviate from the distribution of the normal samples as anomalies.

Autoencoders learn the hidden patterns of the training data by encoding them to a latent space (bottleneck) and reconstructing the embedded data using a decoder. If the input data $x$ is passed through an AE model, the reconstructed output $\hat{x}$ is generated by learning the following representation:
\vspace*{0mm}
\begin{equation}\label{AE_representation}
{\hat{x}(W,b,x)} =\sigma(W^{(n)} \sigma(W^{(n-1)}\sigma{(W^{(n-2)}...})+b^{(n-1)}) + b^{(n)})
\end{equation} 
where $W$ and $b$ represent the AE weight and bias vectors, $n$ is the number of hidden layers, and $\sigma$ represents the non-linear activation function. By learning this representation, the primary objective of the training procedure is to generate the reconstructed data as similar as possible to the normal training data by minimizing the following loss function
\begin{equation}\label{AE_loss}
\mathcal{L_{AE}}\left( W, b, x \right) =\frac{1}{m}\sum_{k=1}^{m}\left| \hat{x}(W,b,x^k)-x^{k} \right|^{2} +\lambda \|W\|^2
\end{equation}
where $m$ is the training data size and $\lambda$ is a hyper-parameter preventing the weights from overgrowing. The second term is the regularization to decrease the magnitudes of the weights.

To extract the hidden patterns of the time series data, we use a convolutional neural network (CNN) in the structure of the encoder and decoder. The CNN model consists of a non-linear activation function and a convolution operator that implements a dot product between the input data and a filter described as:
\begin{equation}\label{CNN}
y^{l+1} = \sigma (W^l * x^l + b^l)
\end{equation}
where $y^{l+1}$ is the output of the ($l+1$)-th layer, $W^l$ and $b^l$ represent weights and biases of the $l$-th layer and $*$ is the convolution operator.
\vspace*{-1mm}

\begin{figure*}
    \hspace{0mm}
    \centering
    \captionsetup{justification=centering}
    \includegraphics[clip,trim=1.7cm 9.0cm 1.1cm 1cm, width=0.99\textwidth]{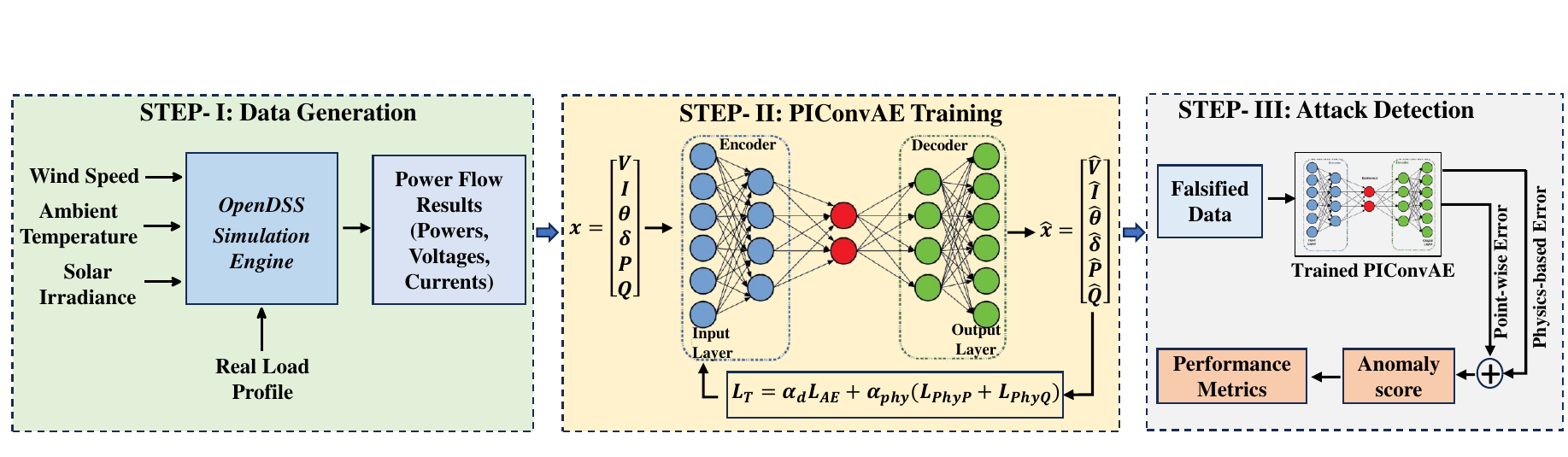}
    \caption{\small Overview of the training and detection of the developed PIConvAE model.}
    \label{fig:PIConvAE_framework}
\end{figure*}
\subsection{Physics-informed Convolutional Autoencoder (PIConvAE)}\label{PIConvAE}
The physics-based principles can be integrated into NN models by various strategies \cite{PIML, PINN}. In this work, an additional loss term is added to the ConvAE loss function to constrain the reconstructed outputs to satisfy the physics-driven laws. For a series of measurements from a measurement unit (vector $x$ in (\ref{input})), the reconstructed outputs should satisfy Kirchhoff’s law at
each node at each time slot $t$. This ensures the physics-based relationships between measurements for power flow calculations. Following this principle, the physics-driven loss functions for the reconstructed measurements at bus $i$ are defined as
\begin{equation}\label{PI_loss_P}
\mathcal{L}_{Phy_P}\left(\hat{x} \right) =\frac{1}{m}\sum_{k=1}^{m}\left| \hat{P}_i^k-\hat{V}_i^k \hat{I}_i^k \cos{(\hat{\theta}_i^k-\hat{\delta}_i^k)} \right|^{2}
\end{equation}
\begin{equation}\label{PI_loss_Q}
\mathcal{L}_{Phy_Q}\left(\hat{x} \right) =\frac{1}{m}\sum_{k=1}^{m}\left| \hat{Q}_i^k-\hat{V}_i^k \hat{I}_i^k \sin{(\hat{\theta}_i^k-\hat{\delta}_i^k)} \right|^{2}
\end{equation}
where $\mathcal{L}_{Phy_P}$ and $\mathcal{L}_{Phy_Q}$ represent the deviations of the reconstructed outputs from active and reactive powers in the Kirchhoff's law, respectively.
The total function for training the PIConvAE model is created by combining (\ref{AE_loss}), (\ref{PI_loss_P}), and (\ref{PI_loss_Q}) as follows 
\begin{equation}\label{loss_T}
\mathcal{L}_{T}\left(W,b,x,\hat{x} \right) = \alpha_d \mathcal{L_{AE}} + \alpha_{phy} (\mathcal{L}_{Phy_P}+\mathcal{L}_{Phy_Q})
\end{equation}
where $\alpha_d$ and $\alpha_{phy}$ are data-driven and physics-based loss coefficients to make a balance between the loss terms. Fig. \ref{fig:PIConvAE_framework} shows the data generation, training, and detection framework of the proposed PIConvAE model. The pseudo-code for the PIConvAE model is shown in Algorithm 1. In this study, we set $\alpha_d =\alpha_{phy}=1$ as it produces better results.

\begin{algorithm}[htbp]\label{alg1}
  \SetAlgoLined
  \SetKwData{Left}{left}
  \SetKwData{This}{this}
  \SetKwData{Up}{up}
  \SetKwFunction{Union}{Union}
  \SetKwFunction{FindCompress}{FindCompress}
  \SetKwInOut{Input}{Input Data}
  \SetKwInOut{Output}{Required}
  \SetKwInOut{Initial}{Initialization}

  \Input{Training, Validation, and Test Data}
  \Output{$Ep$, Number of epochs\\$b$, batch size\\$N_{w}$, window size\\$\alpha$, learning rate}
  \Initial{Weights $W$, $b$}
  
  \BlankLine
  \textbf{Training and Validation:}\\
  \For{each $Ep$}{
    Sample {$(x_i^{1...N_w})_{i=1}^b$} from training data\\
    Sample {$(v_i^{1...N_w})_{i=1}^b$} from validation data
    
    \textbf{Encoder:}\\
    Fix weights $W_{\mathcal{D}}$\\
    $ g_{w_\mathcal{E}}=\nabla_{w_\mathcal{E}}[\alpha_d \mathcal{L_{AE}}\left( W, b, x \right)+ \alpha_{phy} (\mathcal{L}_{Phy_P}(\hat{x})+\mathcal{L}_{Phy_Q}(\hat{x}))]$
    
    \textbf{Decoder:}\\
    Fix weights $W_{\mathcal{E}}$\\
    $ g_{w_\mathcal{D}}=\nabla_{w_\mathcal{D}}[\alpha_d \mathcal{L_{AE}}\left( W, b, x \right)+ \alpha_{phy} (\mathcal{L}_{Phy_P}(\hat{x})+ \mathcal{L}_{Phy_Q}(\hat{x}))]$
    
    Compute training and validation losses
  }
  
  \textbf{Testing:}
  
  $X_{test}$=$(x_n^{1...N_w})_{n=1}^K$
  
  \For{$n=1,...,K$}{
    $\hat{x}^{n}=\mathcal{D}{(\mathcal{E}(x^n))}$
    
    \For{each $l$}{
      ${a}^{n}\left(x^{n},\hat{x}^{n} \right)= |x^{n}-\hat{x}^{n}| + {a}_{phy}^{n}$
    }
  }
  \vspace{-1mm}
  \caption{\small PIConvAE Training, Validation, and Testing}
  \vspace{-0mm}
\end{algorithm}
 \vspace*{-\baselineskip}

\vspace{4mm}
\section{Experimental Results} \label{Results}
The performance of the proposed PIConvAE model is evaluated on the IEEE 13-bus and 123-bus systems. Training, validation, and test data are generated using OpenDSS software. To simulate these systems with DER resources, PV and wind generation units are installed at different nodes to simulate the randomness and uncertainty of the real-world cases. Solar and wind generation data of San Diego city including ambient temperature, solar irradiance, and wind speed for the first week of 2021 are employed \cite{NREL}. Historical load data of San Diego Gas and Electric (SDGE) is used to create load data profiles for both systems. In addition, measurement units are installed at various locations to monitor the measurements. The monitor and DER placements are summarized in Table \ref{tbl:System_Details}. The total generated data has 10080 data points where 75\% is used for training, 10\% for validation, and 15\% for testing. Data samples are normalized using the MinMax function and are divided into subsequences with window size 16 and step size 1 before feeding to the PIConvAE model. Attacks are implemented on the current magnitude at bus 671 in IEEE 13-bus and on the voltage magnitude at bus 35 in IEEE 123-bus systems.  All the simulations are implemented with Tensorflow in Python on a computer with Nvidia RTX3070 8GB and a 2.1 GHz Intel Core i7 CPU with 32 GB of RAM.
\begin{table}[b]
    \centering
    \vspace*{-3mm}
    \caption{\small  DGs and Monitors Locations in the Test Systems.}
    \vspace*{-2mm}
      \includegraphics[clip,trim=6.0cm 5.3cm 0.7cm 5.23cm, width=1.25\linewidth]{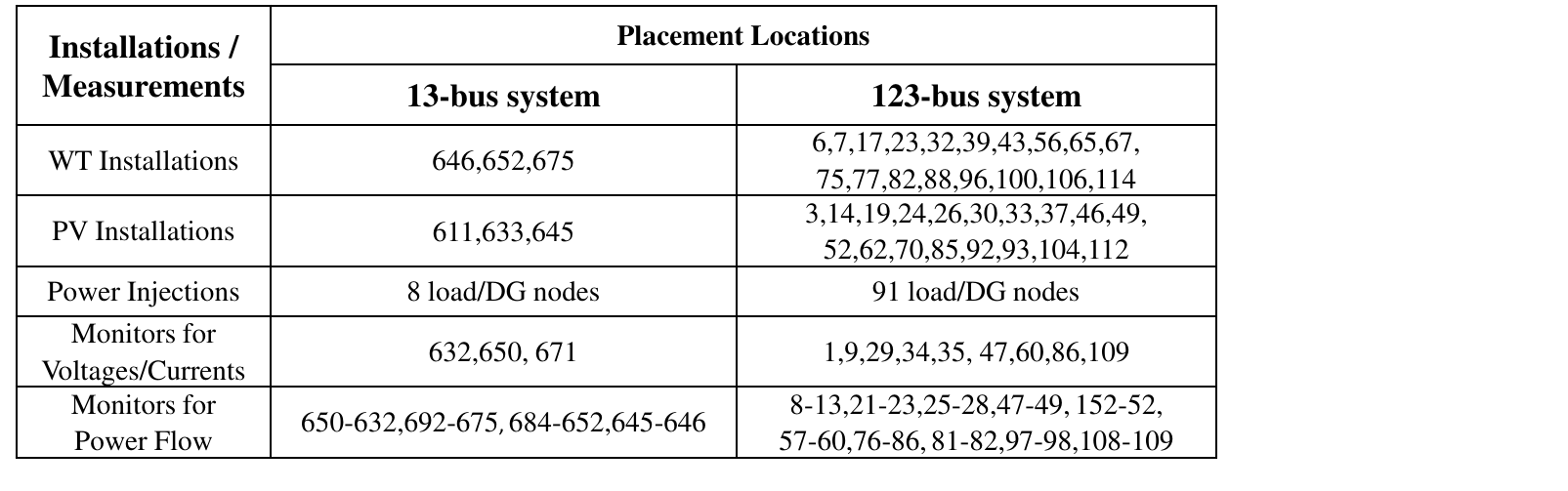}
 
 \label{tbl:System_Details}
\end{table}
\subsection{PIConvAE Settings}\label{parameters}
We employ a two-layer convolutional NN in the encoder design with 64 and 32 filters, with kernel sizes of 5 and 3, respectively. These layers are followed by a fully connected (FC) layer. The decoder has a reverse structure and an FC layer after the second convolutional layer. The activation functions are LeakyReLU and dropout with a rate of 0.2 is applied to the convolutional layers. Batch normalization is applied to all layers and a decaying learning rate with an initial rate of 0.001 is used.
% \begin{figure}[h]
%     \hspace{0mm}
%     \centering
%     % \captionsetup{justification=centering}
%     %\includegraphics[width=1\textwidth]{123.pdf}
%     \includegraphics[clip,trim=9cm 6cm -5cm 2cm, width=0.95\textwidth]{Images/Modified_13-bus_System.pdf}
%     \caption{\small Modified IEEE 13-bus system}
%     \label{fig:13-bus_System}
% \end{figure}
\vspace{-1mm}
\subsection{Performance Metrics}\label{Performance_metrics}
In the testing stage, both reconstruction and physics-based errors are employed to find the anomaly score $a$ at each time step $n$ for measurement $l \in \{V,I,\theta,\delta,P,Q\}$ as follows.
\vspace{0mm}
\begin{equation}
\begin{split}
\label{loss_T}
{a}^{l,n}\left(x^{l,n},\hat{x}^{l,n} \right) &= |x^{l,n}-\hat{x}^{l,n}| + {a}_{phy}^{l,n}
\end{split}
\end{equation}
% \begin{equation}
% \begin{split}
% \label{loss_T}
% {a}^{l,n}\left(x^{l,n},\hat{x}^{l,n} \right) &= |x^{l,n}-\hat{x}^{l,n}| + |P^n-V^n I^n \cos(\theta^n-\delta^n)|^2 \\
% &\quad + |Q^n-V^n I^n \sin(\theta^n-\delta^n)|^2
% \end{split}
% \end{equation}
The first term is the point-wise error and ${a}_{phy}^{l,n}$ represents the physics-based mismatch in the test data. This physics-driven error is defined for each measurement based on Kirchhoff's law. For instance, ${a}_{phy}^{l,n}$ for voltage measurements are defined as 
\vspace{-2mm}
\begin{align}
% \begin{split}
\label{loss_T}
{a}_{phy}^{V,n} &= |V^n-\frac{P^n}{I^n \cos(\theta^n-\delta^n)}|^2
+ |V^n-\frac{Q^n}{I^n \sin(\theta^n-\delta^n)}|^2
\vspace{+1mm}
% \end{split}
\end{align}

To evaluate the performance of the proposed model under attack scenarios, true positive (TP), true negative (TN), false positive (FP), and false negative (FN) are used to find accuracy (ACC=(TP+TN)/(TP+TN+FP+FN), precision (Prec=TP/(TP+FP)), recall (Rec=TP/(TP+FN)), and F-1 score (F1=2$\times$(Prec$\times$ Rec)/(Prec+Rec)).

\subsection{Training and Detection Results}
\begin{figure}[t!]
    \begin{subfigure}{.25\textwidth}
        \centering
        \includegraphics[width=\linewidth]{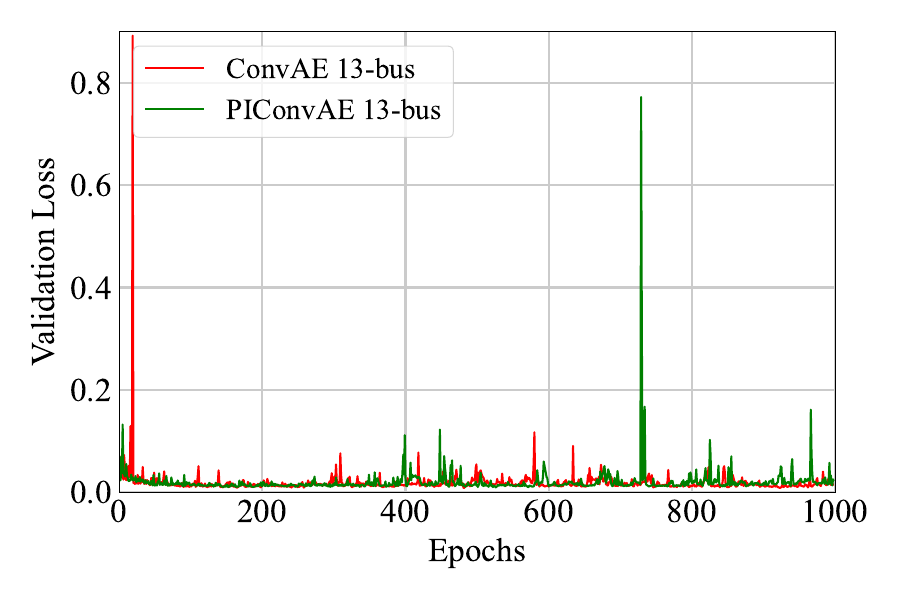}
        \caption{\small Validation loss}
        \label{val_loss_13}
    \end{subfigure}%
    \begin{subfigure}{.25\textwidth}
        \centering
        \includegraphics[width=\linewidth]{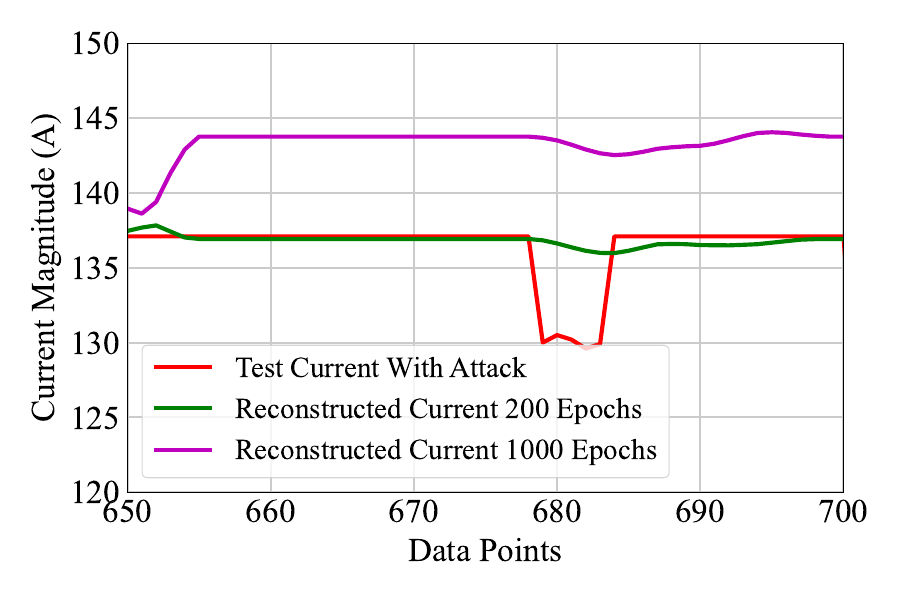}
        \caption{\small Current magnitude}
        \label{Recon_13}
    \end{subfigure}
    
    \begin{subfigure}{.25\textwidth}
        \centering
        \includegraphics[width=\linewidth]{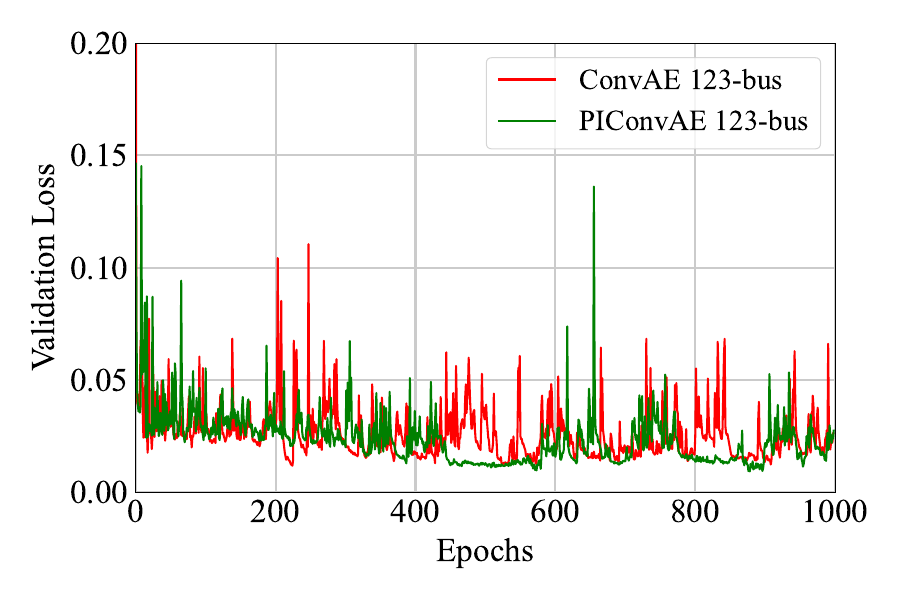}
        \caption{\small Validation loss}
        \label{val_loss_123}
    \end{subfigure}%
    \begin{subfigure}{.25\textwidth}
        \centering
        \includegraphics[width=\linewidth]{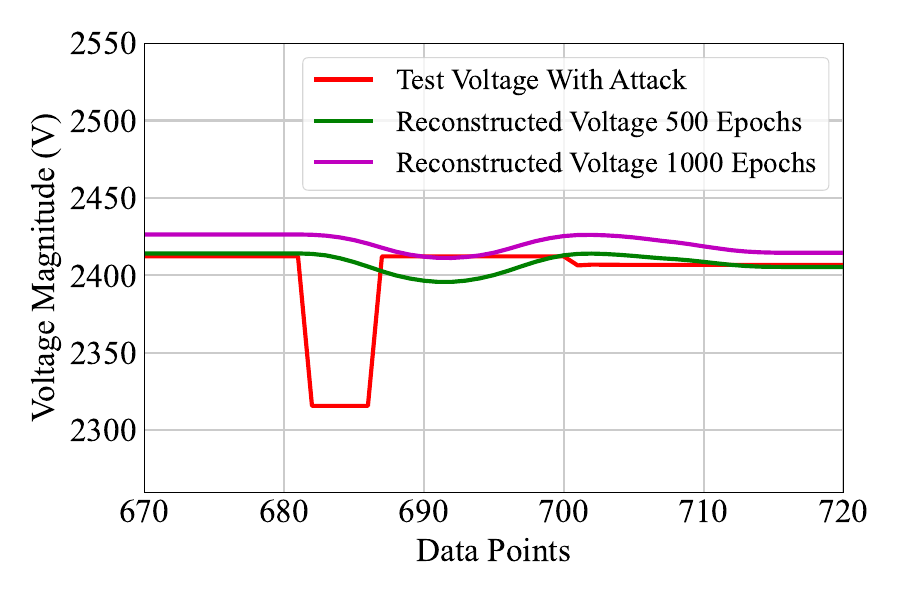}
        \caption{\small Voltage magnitude}
        \label{Recon_123}
    \end{subfigure}

    \caption{\small Simulation results for IEEE 13-bus (a and b), and IEEE 123-bus (c and d).}
    \vspace{-3mm}
    \label{fig:time}
\end{figure}

The proposed PIConvAE model offers advantages such as faster training time and higher detection results. Fig. \ref{val_loss_13} compares the validation loss of the proposed PIConvAE and ConvAE models for the IEEE 13-bus system. As shown in this figure, the loss of the PIConvAE model increases after 200 epochs indicating that the model will be overfit after it while the ConvAE model still requires additional training. This phenomenon of early stop in the training of PIConvAE is clearly illustrated in Fig. \ref{Recon_13} where more training leads to a reduction in the model performance. In other words, the PIConvAE model needs 200 epochs to reconstruct the data whereas the ConvAE model is trained for 1000 epochs. Figures \ref{val_loss_123} and \ref{Recon_123} show the validation loss and reconstruction outputs of the PIConvAE model for the IEEE 123-bus system. As shown in Fig. \ref{val_loss_123}, the validation loss of the PIConvAE model reaches a plateau in around 500 epochs and increases afterward. The performance of the model will degrade when the number of training epochs increases to 1000 as shown in Fig. \ref{Recon_123}. Table \ref{tbl:Time_comparison} also represents the training time of the PIConvAE and ConvAE models for both systems noting that the integration of physics-based equations into the loss function significantly reduces the computational time of PIConvAE compared to the data-driven model.

% \begin{figure}[t!]
%     \subfloat[\small Validation loss]{%
%         \includegraphics[width=.5\linewidth]{Loss_comparison_13.pdf}%
%         % \vspace{-2mm}
%         \label{val_loss_13}%
%     }\hfill
%     \subfloat[\small Current magnitude]{%
%         \includegraphics[width=.5\linewidth]{Im_partial_13.pdf}%
%         % \vspace{-2mm}
%         \label{Recon_13}%
%     }
    
%     \subfloat[\small Validation loss]{%
%         \includegraphics[width=.5\linewidth]{Loss_comparison_123.pdf}%
%         % \vspace{-2mm}
%         \label{val_loss_123}%
%     }\hfill
%     \subfloat[\small Voltage magnitude]{%
%         \includegraphics[width=.5\linewidth]{Vm_partial_123.pdf}%
%         % \vspace{-2mm}
%         \label{Recon_123}%
%     }
    
%     \caption{\small Simulation results for IEEE 13-bus (a and b), and IEEE 123-bus (c and d).}
%     \vspace{-3mm}
%     \label{fig:time}
% \end{figure}
\vspace{-3mm}

\begin{table}[b]
    \centering
    \vspace{-2.5mm}
    \caption{\small  Training time of PIConvAE and ConvAE.}
    \vspace{-3mm}
      \includegraphics[clip,trim=8.0cm 10.3cm -1.7cm 3.91cm, width=1.55\linewidth]{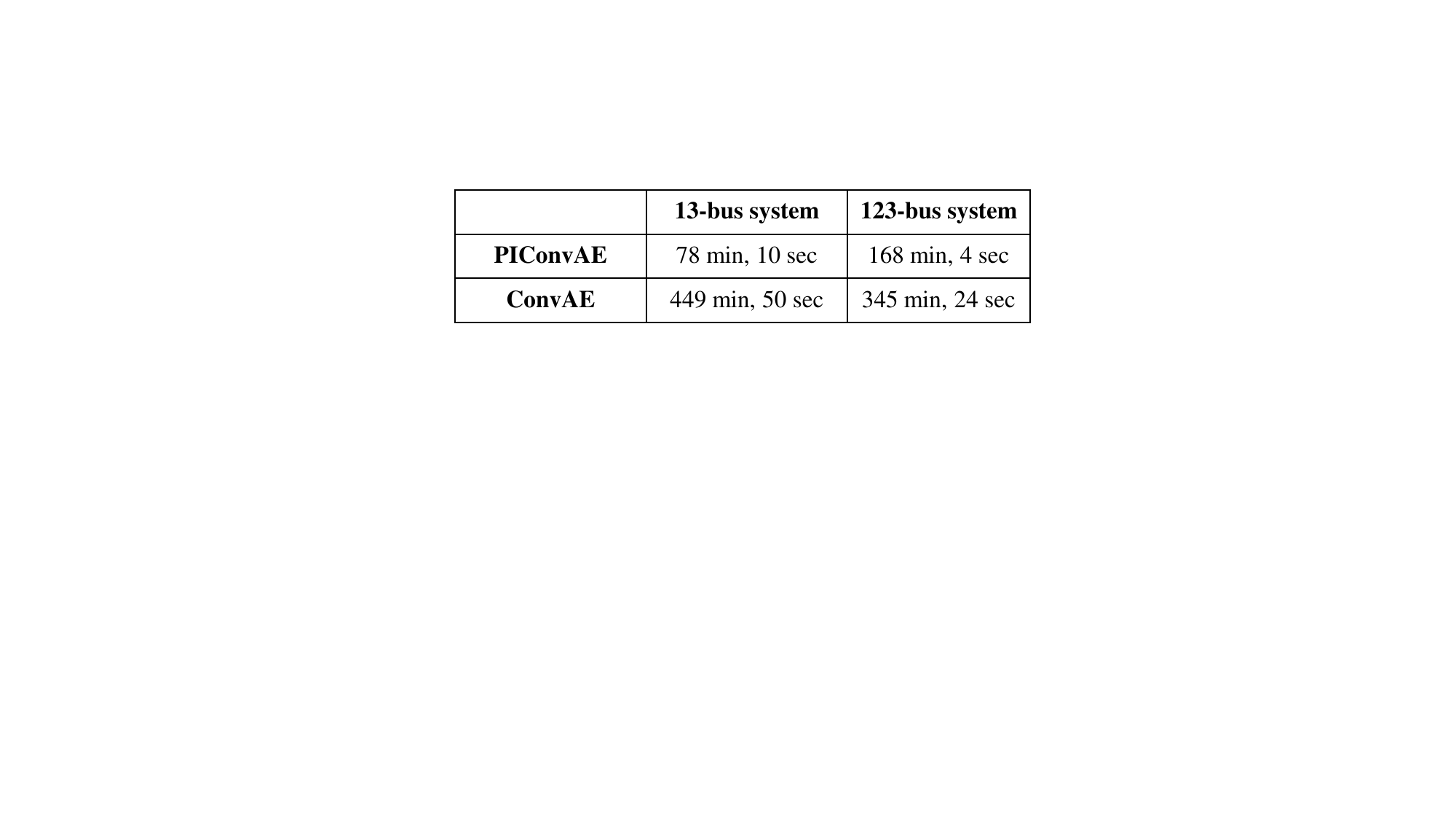}
 \label{tbl:Time_comparison}
  \vspace{-6mm}
\end{table}

\begin{table}[b]
    \centering
    \vspace{-2mm}
    \caption{\small  Performance Comparison for FDIA detection in IEEE 13-bus and 123-bus systems.}
    \vspace{-2mm}
      \includegraphics[clip,trim=6.0cm 7.3cm 0.7cm 4.83cm, width=1.25\linewidth]{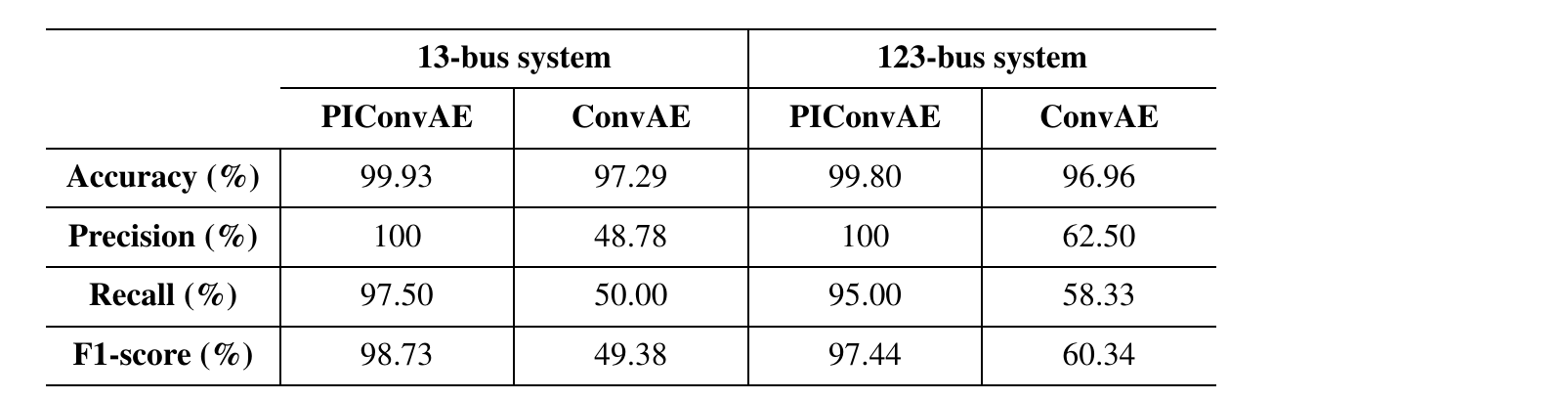}
 \label{tbl:Performance_Metrics}
 \vspace{-6mm}
\end{table}
\vspace{2mm}
In addition to the training time, the PIConvAE model has shown more accurate results for attack detection. Table \ref{tbl:Performance_Metrics} shows the comparison of the detection performance for the two models. As can be seen from this table, in addition to higher accuracy rates, the PIConvAE model has perfect precision for both systems showing the model's capability to correctly detect all the attacked data. It also outperforms the ConvAE model with recall and F-1 score at 97.5\% and 98.73\%, respectively. Although the ConvAE model's performance improves for 123-bus systems, it still lags behind the physics-informed model. The PIConvAE model achieves significantly higher recall and F-1 score compared to the ConvAE model with a recall of 95\% and an F-1 score of 97.44\%.  
The comparison of the physics-informed model with the data-driven model unequivocally demonstrates that by incorporating the physical principles into the NN model, the proposed PIConvAE model achieves higher performances in terms of detection metrics for both systems. This substantiates the outstanding performance of the PIConvAE model in comparison to the data-driven ConvAE model. 
\vspace{-2mm}
\subsection{Evaluation Under Different Attack Magnitudes}
\begin{figure}[t!]
    \centering
    \begin{subfigure}[b]{0.33\textwidth}
    \centering
    \includegraphics[width=\textwidth]{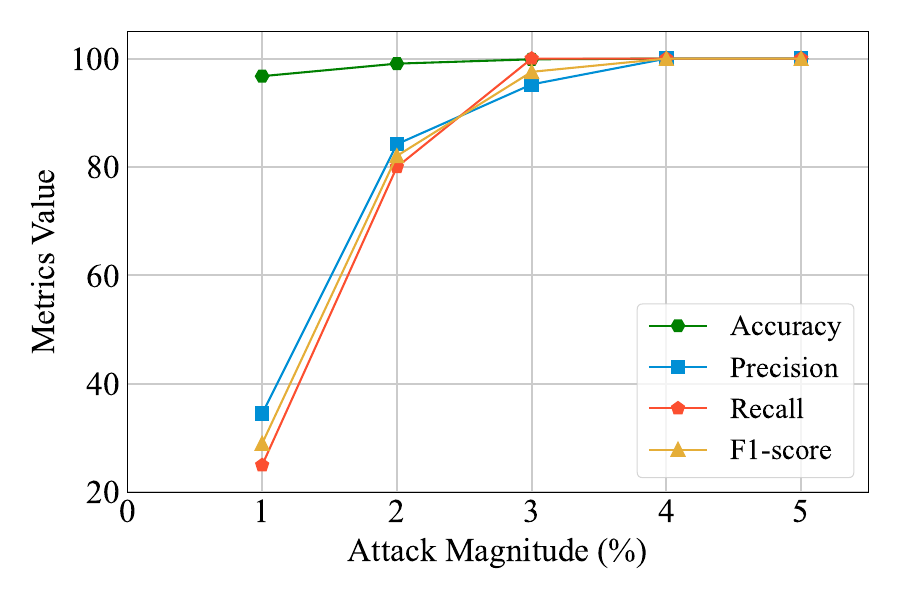}
    \vspace{-8mm}
    \caption{}
    \label{fig:Attack_level_13}
    \end{subfigure}

    \begin{subfigure}[b]{0.33\textwidth}
    \centering
    \includegraphics[width=\textwidth]{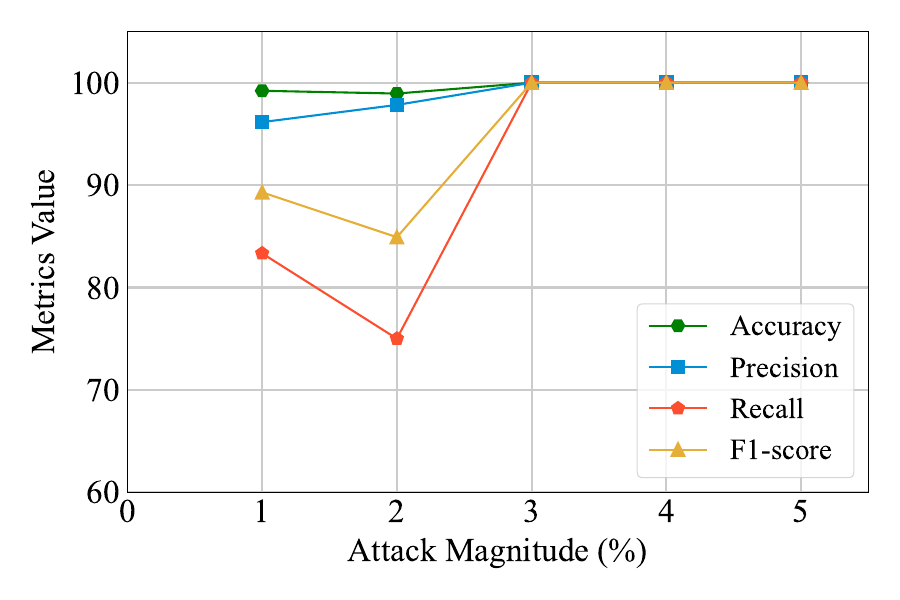}
    \vspace{-8mm}
    \caption{}
    \label{fig:Attack_level_123}
    \end{subfigure}
\caption{\small Sensitivity of the proposed PIConvAE model to different attack magnitudes for (a) IEEE 13-bus system, (b) IEEE 123-bus system.}
\label{fig:Attack_level}
\vspace{-6mm}
\end{figure}

% \begin{figure}[t!]
%     \centering
%     \includegraphics[clip,trim=0.6cm 0.5cm -2.5cm 0cm, width=1.25\linewidth]{Images/Training_loss_13.pdf}
%  \caption{\small  Training loss of Adversarial Autoencoder for IEEE 13-bus power flow dataset.}
%  \label{fig:Training_loss_13}
% \end{figure}
In the previous detection phase, we randomly generated attack magnitudes to be within 5\% of the actual measurements. In this section, the sensitivity of the proposed PIConvAE model against different attack magnitudes is evaluated. Fig. \ref{fig:Attack_level_13} shows performance metrics for the IEEE 13-bus system under different attack magnitudes. As shown in this figure, under very low magnitudes, the model performance is low, however, as the higher magnitudes of attack can be more concerning for the system operation, the proposed model presents remarkable results by the increase in the attack magnitudes. The model achieves superior performance, particularly under attack magnitudes of 3\% and above. The proposed model also exhibits high detection performance for the IEEE 123-bus system. As can be seen in Fig. \ref{fig:Attack_level_123}, the PIConvAE model demonstrates outstanding results for attack detection when the attack magnitude increases.

\section{Conclusion}\label{conclusion}
This work proposed a multivariate PIConvAE model to detect cyber attacks in power distribution grids. The proposed model uses the power flow measurements to embed the physical principles into the loss function of the ConvAE model using Kirchhoff's law. The performance of the proposed PIConvAE model was evaluated on the modified IEEE 13-bus and 123-bus systems. The encoding of physic-based relationships enabled the PIConvAE model to achieve higher detection performance compared to the corresponding data-driven model with faster training time. The detection results on both systems also showed the effectiveness of the proposed model in detecting stealth cyber attacks under different attack magnitudes.

% \begin{figure}
%     \centering
%     \begin{subfigure}[b]{0.45\textwidth}
%     \centering
%     \includegraphics[width=\textwidth]{Images/Test_Vm123_AE_LSTM.pdf}
%     \caption{}
%     \label{fig:AE-LSTM 123-bus}
%     \end{subfigure}

%     \begin{subfigure}[b]{0.45\textwidth}
%     \centering
%     \includegraphics[width=\textwidth]{Images/Test_Vm123_AE_CNN.pdf}
%     \caption{}
%     \label{fig:AE-CNN 123-bus}
%     \end{subfigure}

%         \begin{subfigure}[b]{0.45\textwidth}
%     \centering
%     \includegraphics[width=\textwidth]{Images/Test_Vm123_AE_FC.pdf}
%     \caption{}
%     \label{fig:AE-FC 123-bus}
%     \end{subfigure}
% \caption{\small Outputs of the unsupervised learning models on IEEE 123-bus system test data: (a) AE-LSTM, (b) AE-CNN, (c) AE-FC }
% \label{fig:Test_Vm123_others}
% \end{figure}

\bibliographystyle{ieeetr}
\bibliography{References}
% \end{thebibliography}
%     \bibitem{b1} G. Eason, B. Noble, and I. N. Sneddon, ``On certain integrals of Lipschitz-Hankel type involving products of Bessel functions,'' Phil. Trans. Roy. Soc. London, vol. A247, pp. 529--551, April 1955.
%     \bibitem{b2} J. Clerk Maxwell, A Treatise on Electricity and Magnetism, 3rd ed., vol. 2. Oxford: Clarendon, 1892, pp.68--73.
%     \bibitem{b3} I. S. Jacobs and C. P. Bean, ``Fine particles, thin films and exchange anisotropy,'' in Magnetism, vol. III, G. T. Rado and H. Suhl, Eds. New York: Academic, 1963, pp. 271--350.
%     \bibitem{b4} K. Elissa, ``Title of paper if known,'' unpublished.
%     \bibitem{b5} R. Nicole, ``Title of paper with only first word capitalized,'' J. Name Stand. Abbrev., in press.
%     \bibitem{b6} Y. Yorozu, M. Hirano, K. Oka, and Y. Tagawa, ``Electron spectroscopy studies on magneto-optical media and plastic substrate interface,'' IEEE Transl. J. Magn. Japan, vol. 2, pp. 740--741, August 1987 [Digests 9th Annual Conf. Magnetics Japan, p. 301, 1982].
%     \bibitem{b7} M. Young, The Technical Writer's Handbook. Mill Valley, CA: University Science, 1989.
% \end{thebibliography}
% \vspace{12pt}
% \color{red}
% IEEE conference templates contain guidance text for composing and formatting conference papers. Please ensure that all template text is removed from your conference paper prior to submission to the conference. Failure to remove the template text from your paper may result in your paper not being published.

\end{document}